# Power Allocation in Team Jamming Games in Wireless Ad Hoc Networks

Sourabh Bhattacharya
sbhattac@illinois.edu

Ali Khanafer
khanafe2@illinois.edu

Tamer Başar
basar1@illinois.edu *

## ABSTRACT

In this work, we study the problem of power allocation in teams. Each team consists of two agents who try to split their available power between the tasks of communication and jamming the nodes of the other team. The agents have constraints on their total energy and instantaneous power usage. The cost function is the difference between the rates of erroneously transmitted bits of each team. We model the problem as a zero-sum differential game between the two teams and use *Isaacs'* approach to obtain the necessary conditions for the optimal trajectories. This leads to a continuous-kernel power allocation game among the players. Based on the communications model, we present sufficient conditions on the physical parameters of the agents for the existence of a pure strategy Nash equilibrium (PSNE). Finally, we present simulation results for the case when the agents are holonomic.

## 1. INTRODUCTION

The decentralized nature of wireless ad hoc networks makes them vulnerable to security threats. A prominent example of such threats is jamming: a malicious attack whose objective is to disrupt the communication of the victim network intentionally, causing interference or collision at the receiver side. Jamming attack is a well-studied and active area of research in wireless networks. Unauthorized intrusion of such kind has started a race between the engineers and the hackers; therefore, we have been witnessing a surge of new smart systems aiming to secure modern instrumentation and software from unwanted exogenous attacks.

The problem under consideration in this paper is inspired by recent discoveries of jamming instances in biological species. In a series of playback experiments, researchers have found that resident pairs of Peruvian warbling antbirds sing coordinated duets when responding to rival pairs. But under other circumstances, cooperation breaks down, leading to more complex songs. Specifically, it has been reported that females respond to unpaired sexual rivals by jamming the signals of their own mates, who in turn adjust their signals to avoid the interference [27].

Many defence strategies have been proposed by researchers against jamming in wireless networks. A brief survey of various techniques in jamming relevant to our research is provided in [5]. In the past, networks with multiple attackers have also been considered in the literature. In [15, 16], the authors consider the interaction between a source-destination pair, an eavesdropper, and friendly jammers. The source can buy "jamming power" from the friendly jammers which disguise the eavesdropper. This allows the source to achieve increased secrecy rate. The authors study the problem in the context of a Stackelberg game and show that a trade-off exists between the price announced by the jammers and the resulting performance. A similar problem was tackled in [9] where relay nodes can help the source in the presence of multiple eavesdroppers. The authors propose different relaying schemes and study two design problems: minimizing the transmission power subject to a minimum secrecy rate and maximizing the secrecy rate subject to a total power constraint. Analytical analyses show that relaying yields improved performance when compared to direct transmission in malicious environments. Different from the aforementioned references, our work here considers non-friendly jamming teams, i.e., the security bottleneck considered here is jamming and not eavesdropping. Moreover, to the best of our knowledge, pursuit-evasion strategies for jamming teams were not studied before.

Ad hoc networks consist of mobile energy-constrained nodes. Mobility affects all layers in a network protocol stack including the physical layer as channels become time-varying [13]. Moreover, nodes such as sensors deployed in a field or military vehicles patrolling in remote sites are often equipped with non-rechargeable batteries. Power control plays, therefore, a crucial role in designing robust communications systems. At the physical layer, power control can be used to maximize rate or minimize the transmission error probability, see [18, 3] and the references therein. In addition, in multi-user networks, power control can be used to regulate the interference level at the terminals of other users [24, 28, 10]. Due to the lack of a centralized infrastructure in ad hoc networks, distributed solutions are essential. In this work, similar to [3, 24, 28], we model the power allocation problem as a noncooperative game, thus allowing us to devise a non-centralized solution. As a departure from previous research, however the power control mechanism we propose splits the power budget of each player into two portions: a portion used to communicate with team-mates and a portion used to jam the players of the other team. More impor-

*All three authors are with the Coordinated Science Lab, University of Illinois at Urbana-Champaign, Urbana, IL 61801. This work was supported in part by grants from ARO MURI and AFOSR.

tantly, the objective function is chosen to be the difference between the cumulative bit error rate (BER) of each team; this allows for increased freedom in choosing physical design parameters, besides the power level, such as the size of modulation schemes.

In the past year, there have been reports of predator drones being hacked [14, 20], resulting in intruders gaining access to classified data being transmitted from an aircraft. Motivated from such incidents, we have analyzed various scenarios of evading jamming attacks among autonomous agents. In the case of a *single* jammer trying to intrude the communication link between a transmitter and a receiver, the problem can be formulated as multiplayer (specifically, three player) pursuit-evasion game [17, 2]. In [5], we investigated the problem of finding motion strategies for two unmanned autonomous vehicles (UAVs) to evade jamming in the presence of an aerial intruder. We considered a differential game theoretic approach to compute optimal strategies by a team of UAVs. We formulated the problem as a zero-sum pursuit-evasion game. The cost function was picked as the termination time of the game. We used *Isaacs'* approach to derive the necessary conditions to arrive at the equations governing the saddle-point strategies of the players. In [7], we extended the previous analysis to a team of heterogeneous vehicles containing UAVs and autonomous ground vehicles (AGVs). In [4], we analyzed the problem of multiple jammers intruding the communication network in a formation of UAVs. In [6], we analyzed the problem of connectivity maintenance in multi-agent systems in the presence of a jammer. In this current work, we study a scenario where a *team* of malicious nodes launch a jamming attack on another team, which is capable of jamming as well. We again use differential game theory to study the pursuit-evasion strategies of the teams, which constitute of unmanned decision makers (UDMs).

The main contributions of this paper are as follows. To the best of our knowledge, this is the first work that considers the problem of two teams of mobile autonomous agents jamming each other. Our analysis takes into consideration constraints in energy and power among the agents. Moreover, we relate the problem of optimal power allocation for communication and jamming to the communication model between the agents. Finally, we provide a sufficient condition for existence of an optimal decision strategy among the agents based on the physical parameters of the problems.

The rest of the paper is organized as follows. We formulate the problem in Section 2 and explain the underlying notation. In Section 3, we introduce and solve an associated optimal control problem. The Nash equilibrium properties of the team power control problem are studied in Section 4, and the specific example of systems employing uncoded M-quadrature amplitude modulations (QAM) follows in Section 5. Simulation results are presented in Section 6. We conclude the paper and provide future directions in Section 7. An Appendix at the end includes explicit expressions for some of the variables introduced in the paper.

## 2. PROBLEM FORMULATION

Consider two teams of mobile agents. Each agent is communicating with members of the team it belongs to and at the same time, jamming the communication between members of the other team. We consider a scenario where each team has two members, though at a conceptual level our development applies to higher number of team members as well. Team A is comprised of the two players $\{1^a, 2^a\}$ and Team B is comprised of the two players $\{1^b, 2^b\}$. The agents move on a plane and therefore, have two degrees of freedom $(x, y)$. The dynamics of the players are given by the following equations:

- Team A:
$$\left.\begin{array}{l}\dot{x}_i^a = f_{x_i}^a(\mathbf{x_i^a}, \mathbf{u_i^a}, t) \\ \dot{y}_i^a = f_{y_i}^a(\mathbf{x_i^a}, \mathbf{u_i^a}, t)\end{array}\right\} i \in \{1, 2\} \qquad (1)$$

- Team B:
$$\left.\begin{array}{l}\dot{x}_i^b = f_{x_i}^b(\mathbf{x_i^b}, \mathbf{u_i^b}, t) \\ \dot{y}_i^b = f_{y_i}^b(\mathbf{x_i^b}, \mathbf{u_i^b}, t)\end{array}\right\} i \in \{1, 2\} \qquad (2)$$

In the above equations, $\mathbf{x_i}$ and $\mathbf{u_i}$ denote vectors representing the state and control input of agent $\mathbf{i}$, with the superscript (a or b) identifying the corresponding team. The state space of the entire system is represented by $\mathbf{X} \simeq \mathbb{R}^2 \times \mathbb{R}^2 \times \mathbb{R}^2 \times \mathbb{R}^2$. Moreover, $\mathbf{u_i} \in \mathcal{U}_i \simeq \{\phi : [0, t] \rightarrow \mathcal{A}_i \mid \phi(\cdot) \text{ is measurable}\}$, where $\mathcal{A}_i \subset \mathbb{R}^{p_i}$. $f : \mathbb{R}^2 \times \mathcal{A}_i \times \mathbb{R} \rightarrow \mathbb{R}$ is uniformly continuous, bounded and Lipschitz continuous in $\mathbf{x_i}$ for fixed $\mathbf{u_i}$. Consequently, given a fixed $\mathbf{u_i}(\cdot)$ and an initial point, there exists a unique trajectory solving (1) and (2) [1].

Now, we describe the physical layer communications model in the presence of a jammer which is motivated by [26]. For each transmitter and receiver pair, we assume the following communications model. Given that the transmitter and the receiver are separated by a distance $d$, and the transmitter transmits with constant power $P_T$, the received signal power $P_R$ is given by

$$P_R = \rho P_T d^{-\alpha}, \qquad (3)$$

where $\rho$ depends on the antennas' gains and, according to the free space path loss model, is given by:

$$\rho = \frac{G_T G_R \lambda^2}{(4\pi)^2},$$

where $\lambda$ is the signal's wavelength and $G_T, G_R$ are the transmit and receive antennas' gains, respectively, in the line of sight direction. In real scenarios, $\rho$ is very small in magnitude. For example, using nondirectional antennas and transmitting at 900 MHz, we have $\rho = \frac{1 \cdot 1 \cdot 0.33}{(4\pi)^2} = 6.896 \times 10^{-4}$.

The signal-to-interference ratio (SINR) $s$ is given by

$$s = \frac{P_R}{I + \sigma}, \qquad (4)$$

where $\sigma$ is the ambient noise level. The Bit Error Rate (BER) is given by the following expression:

$$p(t) = g(s), \qquad (5)$$

where $g(\cdot)$ is a decreasing function of $s$. Explicit expressions for $g(\cdot)$ are provided in Section V where we consider the example of M-QAM. Each player uses its power for the following purposes: (1) Communicating with the team-mate, and (2) Jamming the communication of the other team. We assume that $f_a$ and $f_b$ are the frequencies at which Team A and Team B communicate, respectively, and $f_a \neq f_b$.

For an initial position $\mathbf{x}_0 \in \mathbf{X}$, the outcome of the game $\pi$, is given by the following expression:

$$\pi(\mathbf{x}_0, \mathbf{u}_1^a, \mathbf{u}_2^a, \mathbf{u}_1^b, \mathbf{u}_2^b) = \int_0^T \underbrace{[p_1^a(t) + p_2^a(t) - p_1^b(t) - p_2^b(t)]}_{L} dt,$$

where $p_i^a(t)$ and $p_i^b(t)$ are the BERs for agent $i$ in team A and team B, respectively, and $T$ is the time of termination of the game. $p_i$ depends in $s_i$, i.e, the SINR perceived by agent $i$. From (3), $s_i$ depends on the mutual distances between the players. Therefore, we can see that the outcome functional, $\pi$, depends on the state of the players and hence, their control inputs. The outcome functional models the difference in the erroneous communication packets exchanged between the members of the same team during the entire course of the game. The objective of team A is to minimize $\pi$ and the objective of team B is to maximize it.

Let $P_i^a(t)$ and $P_i^b(t)$ denote the instantaneous power levels for communication used by player $i$ in Team A and Team B, respectively. Since the agents are mobile, there are limitations on the amount of energy available to each agent that is dictated by the capacity of the power source carried by each agent. We model this restriction as the following integral constraint for each agent

$$\int_0^T P_i(t)dt \leq E. \quad (6)$$

The game is said to terminate when any one agent runs out of power, that is (6) is violated.

In addition to the energy constraints, there are limitations on the maximum power level of the devices that are used onboard each agent for the purpose of communication. For each player, this constraint is modeled by the following set of inequalities:

$$0 \leq P_i^a(t), P_i^b(t) \leq P_{max}. \quad (7)$$

At every instant, each agent has to decide on the fraction of the power that needs to be allocated for communication and jamming. Table 1 provides a list of decision variables for the players that models this allocation. Each decision variable is a non-negative real number and lies in the interval $[0, 1]$. The decision variables belonging to each row add up to one. The fraction of the total power allocated by the player in row $i$ to the player in column $j$ is given by the first entry in the cell $(i, j)$. This allocated power is used for jamming if the player in column $j$ belongs to the other team; otherwise, it is used to communicate with the agent in the same team. Similarly, the distance between the agent in row $i$ and the agent in column $j$ is given by the second entry in cell $(i, j)$. Since distance is a symmetric quantity, $d^{ij} = d^{ji}$ and $d_{ij} = d_{ji}$.

Figure 1 summarizes the power allocation between the members of the same team as well as between the members of different teams.

In the above game, each agent has to compute the following variables at every instant:

1. The instantaneous control, $\mathbf{u_i}(t)$.

2. The instantaneous power level, $P_i(t)$.

3. All the decision variables present in the row corresponding to the agent in Table 1.

**Table 1: Decision variables and distances among agents.**

|  | $1^b$ | $2^b$ | $1^a$ | $2^a$ |
|---|---|---|---|---|
| $1^a$ | $\gamma_1^1, d_1^1$ | $\gamma_2^1, d_2^1$ |  | $\gamma^{12}, d^{12}$ |
| $2^a$ | $\gamma_1^2, d_1^2$ | $\gamma_2^2, d_2^2$ | $\gamma^{21}, d^{21}$ |  |
| $1^b$ |  | $\delta_{12}, d_{12}$ | $\delta_1^1, d_1^1$ | $\delta_1^2, d_1^2$ |
| $2^b$ | $\delta_{21}, d_{21}$ |  | $\delta_2^1, d_2^1$ | $\delta_2^2, d_2^2$ |

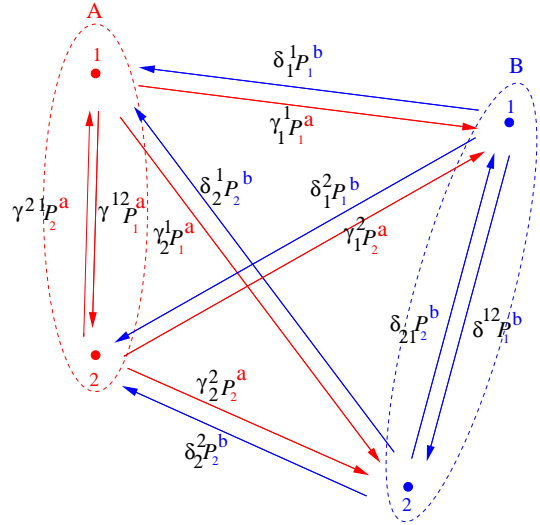

**Figure 1: Power allocation among the agents for communication as well as jamming.**

In the next section, we analyze the problem of computing the optimal controls for each agent.

## 3. OPTIMAL CONTROL PROBLEM

From the problem formulation presented in the previous section, we can conclude that the objective functions of the two teams are in conflict. The tuple $(\mathbf{u}_1^{a*}, \mathbf{u}_2^{a*}, \mathbf{u}_1^{b*}, \mathbf{u}_2^{b*})$ is said to be optimal (or, in pair-wise saddle-point equilibrium) for the players if it satisfies the following conditions:

$$\pi[\mathbf{x}_0, \mathbf{u}_1^{a*}, \mathbf{u}_2^{a*}, \mathbf{u}_1^b, \mathbf{u}_2^b] \leq \pi[\mathbf{x}_0, \mathbf{u}_1^{a*}, \mathbf{u}_2^{a*}, \mathbf{u}_1^{b*}, \mathbf{u}_2^{b*}] \quad (8)$$

$$\pi[\mathbf{x}_0, \mathbf{u}_1^{a*}, \mathbf{u}_2^{a*}, \mathbf{u}_1^{b*}, \mathbf{u}_2^{b*}] \leq \pi[\mathbf{x}_0, \mathbf{u}_1^a, \mathbf{u}_2^a, \mathbf{u}_1^{b*}, \mathbf{u}_2^{b*}] \quad (9)$$

In simple terms, the above equations imply that agents in Team A are solving a joint optimization problem of minimizing the outcome. Similarly, agents in Team B are solving a joint optimization problem of maximizing the outcome. Moreover, the two teams are playing a zero-sum game against one another. In this case, the value

of the game, denoted by the function $J : \mathbf{X} \to \mathbb{R}$, can be defined as follows:

$$J(\mathbf{x}) = \pi[\mathbf{x}_0, \mathbf{u}_1^{a*}, \mathbf{u}_2^{a*}, \mathbf{u}_1^{b*}, \mathbf{u}_2^{b*}] \qquad (10)$$

The value of the game is unique at a point $\mathbf{X}$ in the state-space. An important property satisfied by the value of the game is the *Nash equilibrium* property. The tuple $(\mathbf{u}_1^{a*}, \mathbf{u}_2^{a*}, \mathbf{u}_1^{b*}, \mathbf{u}_2^{b*})$ is said to be in Nash equilibrium if no unilateral deviation in strategy by a player can lead to a better outcome for that player. Hence, there is no motivation for the players to deviate from their equilibrium strategies. In terms of the outcome of the game, the strategies $(\mathbf{u}_1^{a*}, \mathbf{u}_2^{a*}, \mathbf{u}_1^{b*}, \mathbf{u}_2^{b*})$ are in Nash equilibrium (for the 4-player game) if they satisfy the following property:

$$\left. \begin{array}{l} \pi[\mathbf{x}_0, \mathbf{u}_1^{a*}, \mathbf{u}_2^{a*}, \mathbf{u}_1^{b*}, \mathbf{u}_2^{b}] \\ \pi[\mathbf{x}_0, \mathbf{u}_1^{a*}, \mathbf{u}_2^{a*}, \mathbf{u}_1^{b}, \mathbf{u}_2^{b*}] \end{array} \right\} \leq \pi[\mathbf{x}_0, \mathbf{u}_1^{a*}, \mathbf{u}_2^{a*}, \mathbf{u}_1^{b*}, \mathbf{u}_2^{b*}]$$

$$\pi[\mathbf{x}_0, \mathbf{u}_1^{a*}, \mathbf{u}_2^{a*}, \mathbf{u}_1^{b*}, \mathbf{u}_2^{b*}] \leq \left\{ \begin{array}{l} \pi[\mathbf{x}_0, \mathbf{u}_1^{a}, \mathbf{u}_2^{a*}, \mathbf{u}_1^{b*}, \mathbf{u}_2^{b*}] \\ \pi[\mathbf{x}_0, \mathbf{u}_1^{a*}, \mathbf{u}_2^{a}, \mathbf{u}_1^{b*}, \mathbf{u}_2^{b*}] \end{array} \right. \qquad (11)$$

In general, there may be multiple sets of strategies for the players that are in Nash equilibrium. Assuming the existence of a value, as defined in (10), and the existence of a unique Nash equilibrium, we can conclude that the Nash equilibrium concept of person-by-person optimality given in (11) is a necessary condition to be satisfied for the value of the game. Further, obtaining the set of strategies that are in Nash equilibrium yields the optimal strategies for the players. In the following analysis, we assume the aforementioned conditions in order to compute the optimal strategies.

The Hamiltonian of the system is given by the following expression:

$$\begin{aligned} H &= L + \nabla J \cdot f(\mathbf{x}) \\ &= p_1^a(t) + p_2^a(t) - p_1^b(t) - p_2^b(t) + \nabla J \cdot f(\mathbf{x}) \quad (12) \end{aligned}$$

In order to compute the optimal control of the players, we will use the *Isaacs'* conditions [17] which are the following:

1.
$$\left. \begin{array}{l} H[\mathbf{x}_0, \mathbf{u}_1^{a*}, \mathbf{u}_2^{a*}, \mathbf{u}_1^{b*}, \mathbf{u}_2^{b}] \\ H[\mathbf{x}_0, \mathbf{u}_1^{a*}, \mathbf{u}_2^{a*}, \mathbf{u}_1^{b}, \mathbf{u}_2^{b*}] \end{array} \right\} \leq H[\mathbf{x}_0, \mathbf{u}_1^{a*}, \mathbf{u}_2^{a*}, \mathbf{u}_1^{b*}, \mathbf{u}_2^{b*}]$$

$$H[\mathbf{x}_0, \mathbf{u}_1^{a*}, \mathbf{u}_2^{a*}, \mathbf{u}_1^{b*}, \mathbf{u}_2^{b*}] \leq \left\{ \begin{array}{l} H[\mathbf{x}_0, \mathbf{u}_1^{a}, \mathbf{u}_2^{a*}, \mathbf{u}_1^{b*}, \mathbf{u}_2^{b*}] \\ H[\mathbf{x}_0, \mathbf{u}_1^{a*}, \mathbf{u}_2^{a}, \mathbf{u}_1^{b*}, \mathbf{u}_2^{b*}] \end{array} \right.$$

2. $H[\mathbf{x}_0, \mathbf{u}_1^{a*}, \mathbf{u}_2^{a*}, \mathbf{u}_1^{b*}, \mathbf{u}_2^{b*}] = 0$

The agents in Team A want to minimize the Hamiltonian at every instant, and the agents in Team B want to maximize it. The dynamics of the agents are decoupled. Therefore, the Hamiltonian is separable in its controls and, hence, the order of taking the extrema becomes inconsequential. As a consequence, the optimal controls of the players are given by the following expression from Isaacs' first condition.

$$(\mathbf{u}_1^{a*}, \mathbf{u}_2^{a*}, \mathbf{u}_1^{b*}, \mathbf{u}_2^{b*}) = \arg \max_{\mathbf{u}_1^b, \mathbf{u}_2^b} \min_{\mathbf{u}_1^a, \mathbf{u}_2^a} H$$

The optimal control $\mathbf{u}_i$ is obtained by the following expression:

$$\left. \begin{array}{l} \mathbf{u}_i^{a*} = \min_{\mathbf{u}_i^a} \frac{\partial J}{\partial \mathbf{x}_i^a} \cdot f_i^a(\mathbf{x}_i^a, \mathbf{u}_i^a, t) \\ \mathbf{u}_i^{b*} = \max_{\mathbf{u}_i^b} \frac{\partial J}{\partial \mathbf{x}_i^b} \cdot f_i^b(\mathbf{x}_i^b, \mathbf{u}_i^b, t) \end{array} \right\} \quad i = 1, 2$$

Additionally, the gradient of the value function satisfies the *retrogressive path equations* (RPE) given by the following partial differential equation:

$$\frac{\partial \nabla J}{\partial \tau} = \frac{\partial H}{\partial \mathbf{x}},$$

where $\tau$ is the retrograde time or time left for termination.

The RPE leads to the following equations for the players.

**Team A**

$$\mathring{J}_{x_i^a} = \nabla J \cdot \frac{\partial f_{x_i}^a(\mathbf{x})}{\partial x_i^a} + \alpha \sum_{k=1,2} \Big[ \frac{s_k^a g'(s_k^a)(x_j^a - x_i^a)}{(d^{12})^2} + \frac{s_k^b g'(s_k^b) P_{max} \gamma_k^i (x_i^a - x_k^b)}{(d_k^i)^{-(\alpha+2)}} \Big]$$

$$\mathring{J}_{y_i^a} = \nabla J \cdot \frac{\partial f_y^i(\mathbf{x})}{\partial y_i^a} - \alpha \sum_{k=1,2} \Big[ \frac{s_k^a g'(s_k^a)(y_j^a - y_i^a)}{(d^{12})^2} + \frac{s_k^b g'(s_k^b) P_{max} \gamma_k^i (y_i^a - y_k^b)}{(d_k^i)^{-(\alpha+2)}} \Big]$$

**Team B**

$$\mathring{J}_{x_i^b} = \nabla J \cdot \frac{\partial f_{x_i}^b(\mathbf{x})}{\partial x_i^b} - \alpha \sum_{k=1,2} \Big[ \frac{s_k^b g'(s_k^b)(x_j^b - x_i^b)}{(d_{12})^2} + \frac{s_k^a g'(s_k^a) P_{max} \delta_i^k (x_i^b - x_k^a)}{(d_i^k)^{-(\alpha+2)}} \Big]$$

$$\mathring{J}_{y_i^b} = \nabla J \cdot \frac{\partial f_y^i(\mathbf{x})}{\partial y_i^b} + \alpha \sum_{k=1,2} \Big[ \frac{s_k^b g'(s_k^b)(y_j^b - y_i^b)}{(d_{12})^2} + \frac{s_k^a g'(s_k^a) P_{max} \delta_i^k (y_i^b - y_k^a)}{(d_i^k)^{-(\alpha+2)}} \Big]$$

Here, $(\mathring{\cdot})$ denotes derivative with respect to retrograde time. Since termination is only a function of the power of each player, $J$ is independent of the position of the players on the terminal manifold. Therefore, $\nabla J = 0$ at termination. This forms the boundary condition for the RPE.

In the next section, we address the problem of power allocation.

## 4. POWER ALLOCATION

From (4), the SINR received by each agent in terms of the power levels of the other agents as well as their mutual distances is given by the following expressions

$$s_1^a = \frac{P_2^a(t) \gamma^{21} (d^{12})^{-\alpha}}{\frac{\sigma}{\rho} + P_1^b(t) \delta_1^1 (d_1^1)^{-\alpha} + P_2^b(t) \delta_2^1 (d_2^1)^{-\alpha}}$$

$$s_2^a = \frac{P_1^a(t) \gamma^{12} (d^{12})^{-\alpha}}{\frac{\sigma}{\rho} + P_1^b(t) \delta_1^2 (d_1^2)^{-\alpha} + P_2^b(t) \delta_2^2 (d_2^2)^{-\alpha}}$$

$$s_1^b = \frac{P_2^b(t) \delta_{21} (d_{12})^{-\alpha}}{\frac{\sigma}{\rho} + P_1^a(t) \gamma_1^1 (d_1^1)^{-\alpha} + P_2^a(t) \gamma_1^2 (d_1^2)^{-\alpha}}$$

$$s_2^b = \frac{P_1^b(t) \delta_{12} (d_{12})^{-\alpha}}{\frac{\sigma}{\rho} + P_1^a(t) \gamma_2^1 (d_2^1)^{-\alpha} + P_2^a(t) \gamma_2^2 (d_2^2)^{-\alpha}} \qquad (13)$$

From the expression in (12), we can conclude that the power allocation among the agents only affects the term $L$ in the Hamiltonian.

Therefore, Isaacs' first condition leads to the following power allocation problem among the agents.

## 4.1 Team A
The objective of each agent is to minimize $L$.

1. Player 1:
$$\min_{P_1^a,\gamma_1^1,\gamma_2^1,\gamma^{12}} L \Rightarrow \min_{P_1^a,\gamma_1^1,\gamma_2^1,\gamma^{12}} \underbrace{(p_2^a - p_1^b - p_2^b)}_{L_1^a} \quad (14)$$

subject to:
$$P_1^a(t) \leq P_{\max}$$
$$\gamma_1^1 + \gamma_2^1 + \gamma^{12} = 1, \quad \gamma_1^1, \gamma_2^1, \gamma^{12} \geq 0 \Rightarrow \gamma \in \Delta^3$$

2. Player 2:
$$\min_{P_2^a,\gamma_1^2,\gamma_2^2,\gamma^{21}} L \Rightarrow \min_{P_2^a,\gamma_1^2,\gamma_2^2,\gamma^{21}} \underbrace{(p_1^a - p_1^b - p_2^b)}_{L_2^a} \quad (15)$$

subject to:
$$P_2^a(t) \leq P_{\max}$$
$$\gamma_1^2 + \gamma_2^2 + \gamma^{21} = 1, \quad \gamma_1^1, \gamma_2^1, \gamma^{21} \geq 0 \Rightarrow \gamma \in \Delta^3$$

## 4.2 Team B
The objective of each agent is to maximize $L$.

1. Player 1:
$$\max_{P_1^b,\delta_1^1,\delta_2^1,\delta_{12}} L \Rightarrow \max_{P_1^b,\delta_1^1,\delta_2^1,\delta_{12}} \underbrace{(p_1^a + p_2^a - p_2^b)}_{L_1^b} \quad (16)$$

subject to:
$$P_b^1(t) \leq P_{\max}$$
$$\delta_1^1 + \delta_1^2 + \delta_{12} = 1, \quad \delta_1^1, \delta_1^2, \delta_{12} \geq 0$$

2. Player 2:
$$\max_{P_2^b,\delta_2^1,\delta_2^2,\delta_{21}} L \Rightarrow \max_{P_2^b,\delta_2^1,\delta_2^2,\delta_{21}} \underbrace{(p_1^a + p_2^a - p_1^b)}_{L_2^b} \quad (17)$$

subject to:
$$P_b^2(t) \leq P_{\max}$$
$$\delta_2^1 + \delta_2^2 + \delta_{21} = 1, \quad \delta_2^1, \delta_2^2, \delta_{21} \geq 0$$

Since the players do not communicate, they possess information only about their own decision variables. This makes the power allocation problem a continuous kernel non-zero sum game among the players.

THEOREM 1. *The optimal value of the power consumption for each player is $P_{\max}$.*

PROOF. Consider Player $1^a$. The rate $p_a^2$ is a decreasing function of $P_a^1(t)$. Also, $p_1^b$ and $p_2^b$ are increasing functions of $P_a^1(t)$. Therefore, $(p_2^a - p_1^b - p_2^b)$ is a decreasing function of $P_a^1(t)$. Hence the optimal value of $P_a^1(t) = P_{\max}$. Using the same argument for the other players leads to the conclusion that the optimum level of power consumption of every player is $P_{\max}$. □

The next corollary then follows regarding the time horizon of the differential game.

COROLLARY 1. *The entire game terminates in a fixed time $T = \frac{E}{P_{\max}}$ irrespective of the initial position of the agents.*

PROOF. From Theorem 1 and (6), we conclude that the $T = \frac{E}{P_{\max}}$ for all players. All the players consume their entire energy at the same time, i.e., $T$. Therefore, the game ends at time $T$. □

Now we consider the problem of computing the optimal value of the decision variables for the players. In order to do so, we use preexisting results from continuous kernel games that are presented here in Theorem 2 and Theorem 3 and are stated without proof.

THEOREM 2. *[2] An N-person nonzero-sum game in which the finite-dimensional action spaces $U^i$ ($i \in \mathbb{N}$) are compact and cost functionals $J^i$ ($i \in \mathbb{N}$) are continuous on $U^1 \times \cdots \times U^N$ admits a mixed strategy Nash equilibrium (MSNE).*

From the above theorem, we can conclude that the power allocation game has a Nash equilibrium in mixed strategies since the decision variables of each player lie on a simplex which is compact. Moreover, $L$ is a continuous function of the decision variables of all the players. Therefore, the game admits a MSNE. Although, the MSNE has been computed for some games by exploiting some special characteristics in the cost functions, there are no standard techniques to compute MSNE for general continuous-kernel games [22, 2]. Therefore, we search for the conditions under which the power allocation game admits a pure strategy Nash Equilibrium (PSNE).

THEOREM 3. *[2] Let $U$ be a closed, bounded and convex subset of $\mathbb{R}^m$, and for each $i \in \mathbb{N}$ the cost functional $J^i : U \to \mathbb{R}$ be continuous on $U$ and convex in $u^i$ for every $u^j \notin U^j, j \in \mathbb{N}, j \notin i$. Then, the associated N-person nonzero-sum game admits a PSNE.*

The above theorem provides the conditions under which we can guarantee existence of a PSNE. Let us consider the case of agent $1^a$. The expressions for SINR provided in (13) relevant to the optimization problem being solved by $1^a$ can be written in a concise form as shown below:

$$s_2^a = a_1 \gamma^{12}, \quad s_1^b = \frac{b_1}{c_1 + \gamma_1^1}, \quad s_2^b = \frac{d_1}{e_1 + \gamma_2^1},$$

where
$$a_1 = \frac{1}{\frac{\sigma}{P_{max}\rho(d^{12})^{-\alpha}} + \delta_1^2 \left(\frac{d_1^2}{d^{12}}\right)^{-\alpha} + \delta_2^2 \left(\frac{d_2^2}{d^{12}}\right)^{-\alpha}},$$

$$b_1 = \delta_{21} \left(\frac{d_{12}}{d_1^1}\right)^{-\alpha},$$

$$c_1 = \frac{\sigma}{P_{max}\rho(d_1^1)^{-\alpha}} + \gamma_1^2 \left(\frac{d_1^2}{d_1^1}\right)^{-\alpha},$$

$$d_1 = \delta_{12} \left(\frac{d_{12}}{d_2^1}\right)^{-\alpha},$$

$$e_1 = \frac{\sigma}{P_{max}\rho(d_2^1)^{-\alpha}} + \gamma_2^2 \left(\frac{d_2^2}{d_2^1}\right)^{-\alpha}.$$

Note that $a_1, b_1, c_1, d_1$ and $e_1$ are independent of the decision of $1^a$.

THEOREM 4. *The power allocation team game has a unique Nash equilibrium in pure strategies if the following conditions hold for Team A:*

$$g''(s_i^a) > 0, \tag{18}$$

$$g''(s_1^b) + \frac{2}{b_i}(c_i + \gamma_1^i)g'(s_1^b) < 0, \tag{19}$$

$$g''(s_2^b) + \frac{2}{d_i}(e_i + \gamma_2^i)g'(s_2^b) < 0, \tag{20}$$

*and equivalent conditions hold for Team B:*

$$g''(s_i^b) > 0, \tag{21}$$

$$g''(s_1^a) + \frac{2}{l_i}(m_i + \delta_i^1)g'(s_1^a) < 0, \tag{22}$$

$$g''(s_2^a) + \frac{2}{n_i}(o_i + \delta_i^2)g'(s_2^a) < 0, \tag{23}$$

*where* $i \in \{1, 2\}$.

The constants $b_2, c_2, d_2, e_i, l_i, m_i, n_i$, and $o_i$ are obtained by re-writing the SINR expressions as done above; their expressions can be found in the Appendix.

PROOF. Let us consider the case of $1^a$. From Theorem 2, we can conclude that a pure strategy Nash equilibrium exists if $L_1^a$ is convex in its arguments when the decision variables of the other players are fixed. From [8], $L_1^a$ is convex if and only if $\nabla^2 L_1^a > 0$ (For Team B, $L_i^b$ is concave if and only if $\nabla^2 L_i^b < 0$), where the Hessian $\nabla^2 L_1^a$ is given in (24). This constitutes that $g''(s_2^a) > 0$, $g''(s_1^b) + \frac{2}{b_1}(c_1 + \gamma_1^1)g'(s_1^b) < 0$, and $g''(s_2^b) + \frac{2}{d_1}(e_1 + \gamma_2^1)g'(s_2^b) < 0$. The theorem then follows by following similar steps to verify $\nabla^2 L_2^a > 0$, $\nabla^2 L_1^b < 0$, and $\nabla^2 L_2^b < 0$. □

Applying the KKT conditions [19], in addition to the assumptions provided in the theorem that guarantee strict convexity of $L_1^a$, gives us the following equations that need to be satisfied by the globally unique optimal solution ($\bar{\gamma}$):

$$\nabla L_1^a(\bar{\gamma}) + \sum_{i=1}^3 \lambda_i \nabla h_i(\bar{\gamma}) + \eta \nabla h(\bar{\gamma}) = 0$$

$$\left.\begin{array}{c} \lambda_i h_i(\bar{\gamma}) = 0 \\ \lambda_i, \eta \geq 0 \end{array}\right\} \quad i \in \{1, 2, 3\}$$

where

$$\begin{aligned} h_1(\bar{\gamma}) &= -\gamma^{12} \leq 0 \\ h_2(\bar{\gamma}) &= -\gamma_1^1 \leq 0 \\ h_3(\bar{\gamma}) &= -\gamma_2^1 \leq 0 \\ h(\bar{\gamma}) &= \gamma^{12} + \gamma_1^1 + \gamma_2^1 - 1 = 0 \end{aligned}$$

Now, we present the necessary and sufficient conditions for the solution to the optimization problem for the agents. Let us consider the case of $1^a$. The assumptions in Theorem 4 regarding strict convexity of $L_1^a$ render the KKT conditions to be necessary as well as sufficient conditions for the unique global minimum.

To this end, we obtain:

$$\nabla L_1^a = \begin{bmatrix} a_1 g'(s_2^a) \\ \frac{b_1 g'(s_1^b)}{(c_1 + \gamma_1^1)^2} \\ \frac{b_1 g'(s_2^b)}{(c_1 + \gamma_2^1)^2} \end{bmatrix}, \nabla h(\bar{\gamma}) = \begin{bmatrix} 1 \\ 1 \\ 1 \end{bmatrix}$$

$$\nabla h_1(\bar{\gamma}) = \begin{bmatrix} 1 \\ 0 \\ 0 \end{bmatrix}, \nabla h_2(\bar{\gamma}) = \begin{bmatrix} 0 \\ 1 \\ 0 \end{bmatrix}, \nabla h_3(\bar{\gamma}) = \begin{bmatrix} 0 \\ 0 \\ 1 \end{bmatrix}$$

Since $\gamma \in \Delta^3$, at most three of the constraints can be active at any given point. Hence, the gradient of the constraints at any feasible point are always linearly independent.

If two of the three constraints among $\{h_1, h_2, h_3\}$ are active, then $\bar{\gamma}$ has a unique solution that is given by the vertex of the simplex that satisfies the two constraints. If only one of the constraints among $\{h_1, h_2, h_3\}$ is active, then we have the following cases depending on the active constraint

1. $h_1(\bar{\gamma}^1) = 0$: $\bar{\gamma}^1 = (0, \gamma_1^{1*}, 1 - \gamma_1^{1*})$ satisfies the following equations

$$g'(s_2^b)\frac{d_1}{[e_1 + (1 - \gamma_1^{1*})]^2} = g'(s_1^b)\frac{b_1}{[c_1 + \gamma_1^{1*}]^2} \tag{25}$$

2. $h_2(\bar{\gamma}^2) = 0$: $\bar{\gamma}^2 = (1 - \gamma_2^{1*}, 0, \gamma_2^{1*})$ satisfies the following equations

$$a_1 g'(s_2^a) = \frac{d_1 g'(s_2^b)}{(e_1 + \gamma_2^{1*})^2} \tag{26}$$

3. $h_3(\bar{\gamma}^3) = 0$: $\bar{\gamma}^3 = (1 - \gamma_1^{1*}, \gamma_1^{1*}, 0)$ satisfies the following equations

$$a_1 g'(s_2^a) = \frac{b_1 g'(s_1^b)}{(c_1 + \gamma_1^{1*})^2} \tag{27}$$

If none of the inequality constraints are active, then

$$\bar{\gamma}^4 = (\underbrace{1 - \gamma_1^{1*} - \gamma_2^{1*}}_{\gamma^{12*}}, \gamma_1^{1*}, \gamma_2^{1*}),$$

is the solution to the following equations:

$$a_1 g'(s_2^a) - \frac{b_1}{[c_1 + \gamma_1^{1*}]^2}g'(s_1^b) = 0$$

$$a_1 g'(s_2^a) - \frac{d_1}{[e_1 + \gamma_2^{1*}]^2}g'(s_2^b) = 0 \tag{28}$$

Here, $\bar{\gamma}$ lies in the set $\{(1, 0, 0), (0, 1, 0), (0, 0, 1), \bar{\gamma}^1, \bar{\gamma}^2, \bar{\gamma}^3, \bar{\gamma}^4\}$.

An important point to note is that $a_1, b_1, c_1, d_1$ and $e_1$ depend on the decision of the other players. Therefore, the computation of the decision variables depend on the value of the decision variables of the rest of the players. A possible way to deal with this problem is to use iterative schemes for computation of strategies. [2] provides some insights into the efficacy of such schemes from the point of view of convergence and stability. In this work, we assume that each agent has enough computational power so as to complete these iterations in a negligible amount of time compared to the total horizon of the game.

$$\nabla^2 L_1^a = \begin{bmatrix} a_1^2 g''(s_2^a) & 0 & 0 \\ 0 & -\frac{b_1^2}{(c_1+\gamma_1^1)^4}[g''(s_1^b) + \frac{2}{b_1}(c_1+\gamma_1^1)g'(s_1^b)] & 0 \\ 0 & 0 & -\frac{d_1^2}{(e_1+\gamma_2^1)^4}[g''(s_2^b) + \frac{2}{d_1}(e_1+\gamma_2^1)g'(s_2^b)] \end{bmatrix} \quad (24)$$

In the next section, we express the conditions for the existence of PSNE in terms of limitations imposed by the physical communications layer.

## 5. EXISTENCE OF PSNE UNDER M-QAM MODULATION SCHEMES

The bit error rate (BER) depends on the SINR, the modulation scheme, and the error control coding scheme utilized. Communications literature contains closed-form expressions and tight bounds that can be used to calculate $g(s)$ when the noise and interference are assumed to be Gaussian [12]. For example, using uncoded M-QAM, where Gray encoding is used to map the bits into the symbols of the constellation, the BER can be approximated by [23]

$$g(s) \approx \frac{\zeta}{\log(M)} \mathcal{Q}\left(\sqrt{\beta s}\right), \quad (29)$$

where $\zeta = 4(1 - 1/\sqrt{M})$, $\beta = 3/(M-1)$, and $\mathcal{Q}(.)$ is the tail probability of the standard Gaussian distribution which can be expressed in terms of the error function:

$$\mathcal{Q}(x) = \frac{1}{2} - \frac{1}{2}\text{erf}\left(\frac{x}{\sqrt{2}}\right).$$

The conditions of Theorem 4 depend primarily on the employed modulation and coding schemes.

THEOREM 5. *When all players employ uncoded M-QAM modulation schemes, the power allocation team game has a unique PSNE solution if the following condition is satisfied:*

$$\beta \rho P_{max} \left(\min\{d^{12}, d_{12}\}\right)^{-\alpha} < 3\sigma. \quad (30)$$

PROOF. The conditions of Theorem 4 need to be satisfied for a unique pure strategies solution to exist. We first verify those conditions for Player $1^a$ when uncoded M-QAM modulations are used. To this end, we differentiate (29) in to obtain:

$$g'(s) = -\frac{\zeta\sqrt{\beta}}{2\log(M)\sqrt{2\pi s}} \exp\left(-\frac{\beta}{2}s\right),$$

$$g''(s) = \frac{\zeta\sqrt{\beta}(1+\beta\sqrt{s^2})}{4\log(M)\sqrt{2\pi s^3}} \exp\left(-\frac{\beta}{2}s\right). \quad (31)$$

From (31), we conclude that condition (18) holds for any value of $a_1$ and $\gamma^{12}$. Condition (19) holds given that

$$c_1 > \frac{\beta}{3}b_1 - \gamma_1^1,$$

or equivalently

$$c_1 > \frac{\beta}{3}b_1,$$

which we can re-write as

$$\frac{\sigma}{P_{max}\rho} + \gamma_1^2(d_1^2)^{-\alpha} > \frac{\beta}{3}\delta_{21}(d_{12})^{-\alpha}, \quad (32)$$

or:

$$\frac{\sigma}{P_{max}\rho} > \frac{\beta}{3}(d_{12})^{-\alpha}. \quad (33)$$

By following similar steps, we can show that (33) is sufficient for (20) to hold. In fact, condition (33) is also sufficient for the convexity of $L_2^a$. For Team B, a sufficient condition for the concavity of $L_1^b$ and $L_2^b$ is

$$\frac{\sigma}{P_{max}\rho} > \frac{\beta}{3}(d^{12})^{-\alpha}, \quad (34)$$

which can be derived following similar steps to the above. The theorem follows from (33) and (34). □

Note that the left hand side of inequality (30) depends entirely on physical design parameters; this is of particular importance for design purposes. Moreover, sufficient conditions for Theorem 5 can be expressed in terms of the received SNRs for all players, which could be more insightful from a communication systems prospective. Consider, for example, Player $1^a$, and let $\text{SNR}_y^x = \frac{P_{max}\gamma_y^x\rho(d_y^x)^{-\alpha}}{\sigma}$ and $\text{SNR}_{xy} = \frac{P_{max}\delta_{xy}\rho(d_{yx})^{-\alpha}}{\sigma}$. Expression (32) can then be written as

$$\text{SNR}_{21} < \frac{3}{\beta}(\text{SNR}_1^2 + 1).$$

Similarly, condition (20) holds if

$$\text{SNR}_{12} < \frac{3}{\beta}(\text{SNR}_2^2 + 1).$$

Yet another useful way to interpret condition (30) is regarding it as a minimum rate condition:

$$R > \log\left(1 + \frac{\rho P_{max}\left(\min\{d^{12}, d_{12}\}\right)^{-\alpha}}{\sigma}\right),$$

where $R = \log(M)$.

The specific conditions for Player $1^a$ corresponding to (25)-(27) when M-QAM modulations are utilized are:

$$\left(\frac{s_1^b}{s_2^b}\right)^{\frac{3}{2}} \exp\left(-\frac{\beta}{2}(s_1^b - s_2^b)\right) - \frac{b_1}{d_1} = 0,$$

$$\left(\frac{s_2^b}{s_2^a}\right)^{\frac{1}{2}} \exp\left(-\frac{\beta}{2}(s_2^a - s_2^b)\right) - \frac{a_1 d_1}{(e_1+\gamma_2^1)^2} = 0,$$

$$\left(\frac{s_1^b}{s_2^a}\right)^{\frac{1}{2}} \exp\left(-\frac{\beta}{2}(s_2^a - s_1^b)\right) - \frac{a_1 b_1}{(c_1+\gamma_1^1)^2} = 0.$$

Also, (28) in this case becomes

$$\left(\frac{s_2^b}{s_2^a}\right)^{\frac{1}{2}} \exp\left(-\frac{\beta}{2}(s_2^a - s_2^b)\right) - \frac{a_1 d_!}{(e_1+\gamma_2^1)^2} = 0,$$

$$\left(\frac{s_1^b}{s_2^a}\right)^{\frac{1}{2}} \exp\left(-\frac{\beta}{2}(s_2^a - s_1^b)\right) - \frac{a_1 b_1}{(c_1+\gamma_1^1)^2} = 0.$$

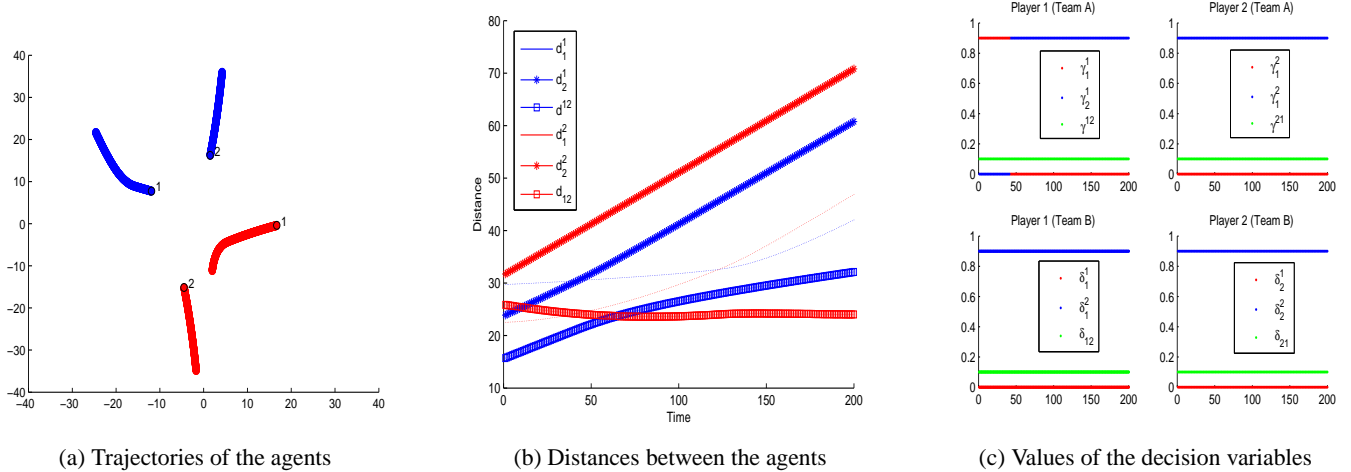

(a) Trajectories of the agents  (b) Distances between the agents  (c) Values of the decision variables

Figure 2:

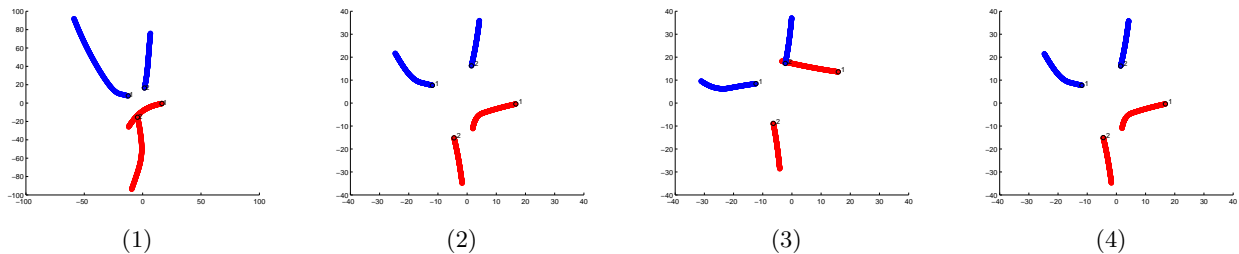

(1)  (2)  (3)  (4)

Figure 3: Trajectories of the agents on the plane

## 6. SIMULATION RESULTS

In this section, we present some simulation results. First, we present simulations for arbitrarily chosen values of maximum power $P_{max}$, frequencies $f_1, f_2$, modulation scheme size $M$, and speeds of the players $u$ satisfying the convexity conditions in the previous section. We then change the value of one parameter, while fixing the rest, and present simulation results for the variants of the original problem.

Figure 2 shows the trajectories of the agents, distances between agents, and the values of the decision variables over 200 time steps. The kinematics of all the agents are given by the following equations

$$\dot{x}_i = u_i \cos \theta_i, \quad \dot{y}_i = u_i \sin \theta_i$$

We used the following values for the parameters in the simulations:

- $P_{\max} = 100$
- $f_1 = 300$ MHz and $f_2 = 100$ MHz
- Typical values of M are 2, 4, 16, 64, and 256. For simulation purposes, we fix the value of $M$ at 2.
- $u_1^a = u_2^a = u_1^b = u_2^b = 1$

Figures 3, 4, and 5 contain four subfigures each repeating one of the above simulations for a variation of the original parameters as follows:

1. Different Speeds: $u_1^a = 5, u_2^a = 1, u_1^b = 3$, and $u_2^b = 4$
2. Different Modulation: $M = 16$
3. Different terminal conditions.
4. Frequency exchange: $f_1 = 100$MHz and $f_2 = 300$MHz

In each case, we present the simulation results when all the parameters are fixed except for the one listed above.

## 7. CONCLUSION AND FUTURE WORK

This paper has studied the power allocation problem for jamming teams. The motion of the teams was modelled using the framework of pursuit-evasion games and the optimal strategies were derived. An underlying static game was used to obtain the optimal power allocation, where the power budget of each user is split between communication and jamming powers. This work focused on the analysis of teams consisting of two players only. Potential future directions include:

- *Computation of Singular Surfaces*: In this work, we have computed the trajectories based on the necessary conditions of optimality imposed by the Isaacs' conditions. In order to complete the construction of the optimal trajectories of the agents, we have to identify the singular surfaces in the state space [2]. This is an interesting future research direction since the construction and nature of the singular surfaces

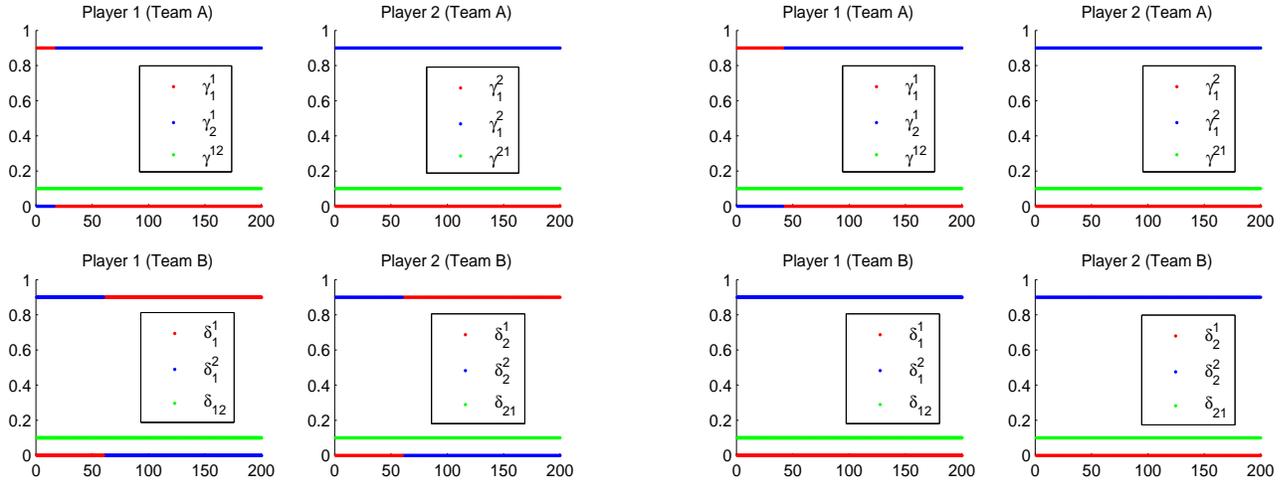

(1) (2)

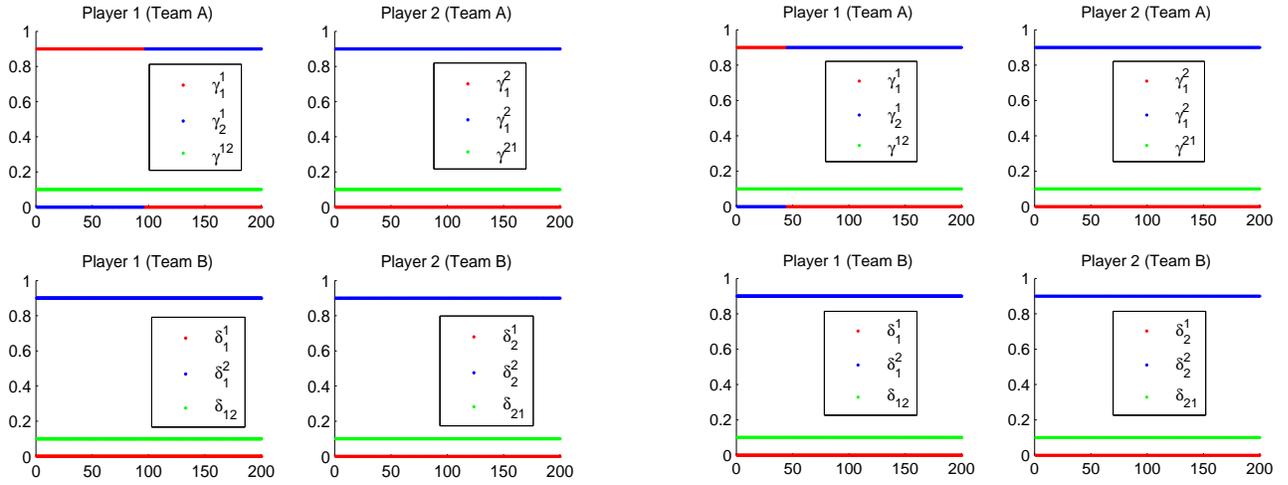

(3) (4)

Figure 4: **Variations of the decision variables as a function of time.**

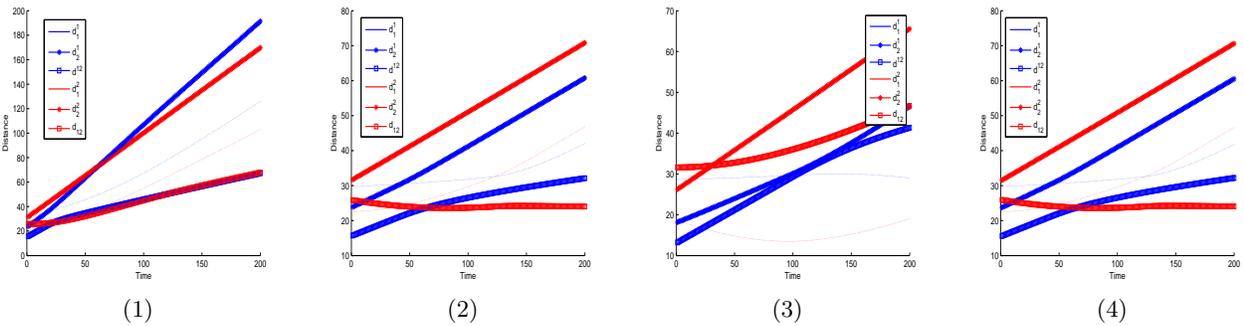

(1) (2) (3) (4)

Figure 5: **Distances among the agents**

would depend on the value of the decision variables obtained from the power allocation game.

- *Computation of MSNE*: As discussed in Section 3, the power allocation games admits a MSNE without any constraints on the underlying communication model. An important future problem is to compute the MSNE for the power allocation game.

- *Scheduling Schemes*: An interesting direction would be exploring scheduling algorithms, similar to the one proposed in [11], in which players take turns in communicating or jamming. For example, the users of a given team that are closest in distance to the the other team could allocate all their resources to jamming, while the other users allocate all their resources to communicating with each other.

- *Power Control*: When multiple users are present, and due to the broadcast nature of wireless systems, networks become interference-limited. The transmission power of one user can impede the links between other nodes due to the interference; hence, it is important to regulate the transmission power of the users in order to, for example, maximize the total capacity of the network.

- *Routing*: Multihop routing improves the total throughput and power efficiency of a network through relaying packets via intermediate nodes to their final destination. Because a portion of the energy of each node has to be allocated to jam the other team, determining the optimal route for transmission becomes a challenge, especially in the presence of mobility. An investigation of routing protocols in the context of games is therefore essential for studying the overall performance of the networks [25].

- *Eavesdropping:* When $f_a = f_b$, another security issue arises as the ADMs of a given team can receive and decode messages intended for internal communications of other teams. To ensure secure communications, each team would need to allocate power to jam the eavesdroppers. In fact, a more general scenario is when adversarial teams consist of active eavesdroppers: malicious nodes that can act as jammers and eavesdroppers [21].

# APPENDIX

The following are the expressions of the quantities appearing in (18)-(23) and in $\nabla^2 L_2^a$, $\nabla^2 L_1^b$, and $\nabla^2 L_2^b$:

$$a_2 = \frac{1}{\frac{\sigma}{P_{max}\rho(d^{12})^{-\alpha}} + \delta_1^1 \left(\frac{d_1^1}{d^{12}}\right)^{-\alpha} + \delta_2^1 \left(\frac{d_2^1}{d^{12}}\right)^{-\alpha}}$$

$$b_2 = \delta_{21} \left(\frac{d_{12}}{d_1^2}\right)^{-\alpha}$$

$$c_2 = \frac{\sigma}{P_{max}\rho(d_1^2)^{-\alpha}} + \gamma_1^1 \left(\frac{d_1^1}{d_1^2}\right)^{-\alpha}$$

$$d_2 = \delta_{12} \left(\frac{d_{12}}{d_2^2}\right)^{-\alpha}$$

$$e_2 = \frac{\sigma}{P_{max}\rho(d_2^2)^{-\alpha}} + \gamma_2^1 \left(\frac{d_2^1}{d_2^2}\right)^{-\alpha}$$

$$k_1 = \frac{1}{\frac{\sigma}{P_{max}\rho(d_{12})^{-\alpha}} + \gamma_2^1 \left(\frac{d_2^1}{d_{12}}\right)^{-\alpha} + \gamma_2^2 \left(\frac{d_2^2}{d_{12}}\right)^{-\alpha}}$$

$$l_1 = \gamma^{21} \left(\frac{d^{12}}{d_1^1}\right)^{-\alpha}$$

$$m_1 = \frac{\sigma}{P_{max}\rho(d_1^1)^{-\alpha}} + \delta_2^1 \left(\frac{d_2^1}{d_1^1}\right)^{-\alpha}$$

$$n_1 = \gamma^{12} \left(\frac{d^{12}}{d_1^2}\right)^{-\alpha}$$

$$o_1 = \frac{\sigma}{P_{max}\rho(d_1^2)^{-\alpha}} + \delta_2^2 \left(\frac{d_2^2}{d_1^2}\right)^{-\alpha}$$

$$k_2 = \frac{1}{\frac{\sigma}{P_{max}\rho(d_{12})^{-\alpha}} + \gamma_1^1 \left(\frac{d_1^1}{d_{12}}\right)^{-\alpha} + \gamma_1^2 \left(\frac{d_1^2}{d_{12}}\right)^{-\alpha}}$$

$$l_2 = \gamma^{21} \left(\frac{d^{12}}{d_2^1}\right)^{-\alpha}$$

$$m_2 = \frac{\sigma}{P_{max}\rho(d_2^1)^{-\alpha}} + \delta_1^1 \left(\frac{d_1^1}{d_2^1}\right)^{-\alpha}$$

$$n_2 = \gamma^{12} \left(\frac{d^{12}}{d_2^2}\right)^{-\alpha}$$

$$o_2 = \frac{\sigma}{P_{max}\rho(d_2^2)^{-\alpha}} + \delta_1^2 \left(\frac{d_1^2}{d_2^2}\right)^{-\alpha}$$